\documentclass[sigconf]{acmart}
\usepackage{natbib}
\usepackage[utf8]{inputenc} 
\usepackage[T1]{fontenc}    
\usepackage{hyperref}       
\usepackage{url}            
\usepackage{booktabs}       
\usepackage{amsfonts}       
\usepackage{nicefrac}       
\usepackage{microtype}      
\usepackage{xcolor}         
\usepackage{indentfirst}
\usepackage{amsmath,graphicx}
\usepackage{amsfonts}  
\usepackage{dsfont}
\usepackage{subfigure}
\usepackage{bm}
\usepackage{algorithm}               
\usepackage{algorithmic}             
\usepackage{multirow}                
\usepackage{color}
\usepackage{multicol}                
\usepackage{setspace}
\usepackage{lipsum}

\title{Edge Graph Neural Networks for Massive MIMO Detection}

	\author{Hongyi Li}
	\affiliation{%
		\institution{State Key Laboratory of ISN\\Xidian University}
		\city{Xi'an}
		\country{China}
		\
	}
		\author{Junxiang Wang}
	\affiliation{%
		\institution{Department of Computer Science\\ Emory University}
		\city{Atlanta}
		\country{USA}
	}
		\author{Yongchao Wang}
	\affiliation{%
		\institution{State Key Laboratory of ISN\\Xidian University}
		\city{Xi'an}
		\country{China}
		\
	}
\begin{document}
%
\newcommand\blfootnote[1]{%
\begingroup
\renewcommand\thefootnote{}\footnote{#1}%
\addtocounter{footnote}{-1}%
\endgroup
}
\begin{abstract}
Massive Multiple-Input Multiple-Out (MIMO) detection is an important problem in modern wireless communication systems. While traditional Belief Propagation (BP) detectors perform poorly on loopy graphs, the recent Graph Neural Networks (GNNs)-based method can overcome the drawbacks of BP and achieve superior performance. Nevertheless, direct use of GNN ignores the importance of edge attributes and suffers from high computation overhead using a fully connected graph structure. In this paper, we propose an efficient  GNN-inspired algorithm, called the Edge Graph Neural Network (EGNN), to detect MIMO signals. We first compute graph edge weights through channel correlation and then leverage the obtained weights as a metric to evaluate the importance of neighbors of each node. Moreover, we design an adaptive Edge Drop (ED) scheme to sparsify the graph such that computational cost can be significantly reduced. Experimental results demonstrate that our proposed EGNN achieves better or comparable performance to popular MIMO detection methods for different modulation schemes and costs the least detection time compared to GNN-based approaches. 
\end{abstract}
\maketitle

\section{Introduction}
\indent Massive Multiple-Input Multiple-Out (MIMO) plays an important role  in modern wireless systems, which applies multiple antennas at both the transmitter and the receiver side to advance spectral and power efficiencies. Since the transmitted symbols are coupled and corrupted at the receiver side, the detection technology is required to recover the transmitted signals, which is referred as MIMO detection.   \citep{albreem2019massive,tan2019improving}.
Though the Maximum-Likelihood (ML) estimation-based MIMO detector can achieve the optimal solution, it is computationally intractable as it requires to enumerate all valid transmitted symbol vectors. To deal with this challenge, many approximation methods have been proposed  \citep{gao2017massive,goldberger2010pseudo,liu2019capacity,som2010improved,tan2019improving,yedidia2003understanding,tiba2021alow, zhang2022effective}. For example,
Belief propagation (BP)-based detectors infer the transmitted symbols by iterative message dispatching in a graphical model without matrix inversion \citep{yedidia2003understanding}.
One of its variants, the Markov Random Field (MRF) detector, models transmitting antennas and channel information as nodes and edges, respectively \citep{tan2019improving}. However, they suffer from several challenges including severe performance degradation on loopy graphs and hand-crafted message-passing functions \citep{goldberger2010pseudo}. Recently, Graph Neural Networks (GNNs) have attracted much attention from machine learning researchers due to their tremendous success on many graph learning tasks such as node classification and link prediction \cite{kipf2016semi}. This is because GNN can learn node  representations effectively via aggregating information from neighboring nodes. Some existing works has applied them into  MIMO detectors such as surrogate messages and lead to better performance compared with BP \citep{scotti2020graph}. However, they still suffer from two challenges: firstly, they still can not fully leverage edge information; secondly, they are computationally expensive due to massive message propagation in GNNs.

\indent In order to address these challenges simultaneously, we propose an efficient GNN-inspired MIMO detection algorithm named Edge Graph Neural Networks (EGNN), which improves performance further by exploiting edge attributes effectively. Specifically, we compute edge weights by channel correlation and treat obtained weights as metrics to evaluate the importance of neighbors. Moreover, we develop an adaptive Edge Drop (ED) scheme to sparsify a graph, in order to reduce the computational cost. Experimental results demonstrate the effectiveness and efficiency of our proposed EGNN compared with state-of-the-art MIMO detection methods.
\section{Background}\label{problem formulation}

\subsection{Problem Formulation}
Consider an uplink multi-user MIMO system with $N_t$ transmit antennas and $N_r$ receive antennas, where each user sends an independent symbol. The equivalent real-valued model of the MIMO system is given by $\mathbf{y} =  \mathbf{Hx} + \mathbf{n}$.
Given the received signal $\mathbf{y} \in \mathbb{R}^{2N_r}$, the variance of the zero-mean Additive White Gaussian Noise (AWGN) $\sigma^2$ and the Rayleigh  channel matrix $\mathbf{H}\in \mathbb{R}^{2N_r \times 2N_t}$, the aim of the MIMO detection is to infer the transmitted symbols $\mathbf{x}=[x_1, x_2, \cdots, x_{2N_t}]^T \in \mathcal{A}^{2N_t}$, where $\mathcal{A}=\{s_1, s_2, \cdots, s_K\}$ is a real-valued finite alphabet set whose size $K$ is the square root of the modulation order.
\subsection{The BP Detection Method}
The Maximum\emph{ A Posteriori} (MAP) solution to the above system  is defined by
\begin{equation}\label{map solution}
\begin{aligned}
\hat{\mathbf{x}}_{MAP} = \mathop {\arg \max }\nolimits_{\mathbf{x}\in \mathcal{A}^{2Nt}} p(\mathbf{x}|\mathbf{y}).
\end{aligned}
\end{equation}
The BP detector factorizes the joint posterior probability function $p(\mathbf{x}|\mathbf{y})$ into a product of local functions that only contain either single or two variables:
{\setlength\abovedisplayskip{1pt}
\setlength\belowdisplayskip{1pt}

\begin{equation}\label{bp model}
\begin{aligned}
p(\mathbf{x}|\mathbf{y}) & \propto  p(\mathbf{x})p\left(\mathbf{y}|\mathbf{x}\right)
  \\
  &= p(\mathbf{x}) \cdot e^{-\frac{1}{2\sigma^2}\|\mathbf{Hx-y}\|^2 }\\
  &
  \propto\left(\prod\nolimits_i\phi_i(x_i)\right)\left(\prod\nolimits_{i< j}\phi_{ij}(x_i,x_j)\right),
\end{aligned}
\end{equation}
}where
$\phi_i(x_i) =   p_i(x_i)\cdot e^{-\frac{1}{2\sigma^2}(\mathbf{h}_i^T\mathbf{h}_i x_i^2 - 2\mathbf{y}^T\mathbf{h}_ix_i)}$ and $
\phi_{ij}(x_i,x_j) = e^{-\frac{1}{\sigma^2}\mathbf{h}_i^T\mathbf{h}_jx_ix_j}$,  $p(\mathbf{x})$ and $p_i(x_i)$ denote the joint prior probability and  the prior of the variable $x_i$, respectively, and $\mathbf{h}_i$ is the $i$-th column of $\mathbf{H}$. In most cases the prior probabilities are assumed uniform over symbols.
We can develop a Markov Random Field (MRF) and achieve the maximum of $p(\mathbf{x}|\mathbf{y})$ by dispatching messages between nodes.
\subsection{GNNs for MIMO Detection}
\indent Based on the MRF graph, previous works  transformed the signal detection problem to a node classification task, which can be solved by GNNs \cite{scotti2020graph,yoon2019inference}. The GNNs consider a graph $\mathrm{G} = \{\mathbf{E}, \mathbf{Z}^{(0)}\}$ as an input at step $0$, in which $\mathbf{Z}^{(0)}=[\mathbf{z}_1^{(0)},\mathbf{z}_2^{(0)}, \cdots, \mathbf{z}_N^{(0)}]$, wherein  $\mathbf{z}_i^{(0)}$ stands for the initial feature of node $v_i$, and $\mathbf{E}=\{\epsilon_{ij}\}_{N \times N}$ is a matrix that each entry $\epsilon_{ij}$  represents the attribute of the directed edge from node $i$ to $j$. GNNs iteratively update vectors by message passing between neighboring nodes for $T$ steps sequentially and finally project vectors to target outputs.

A GNN framework consists of three functions: a message function together with an update function for iterative node update, and a readout function for node classification. At step $t$, the central node $i$ receives messages from all of its neighbors $j$ by the message function: $\mathbf{\bf{m}}_{j \to i}^{(t)} = \mathbf{M}(\mathbf{z}_{i}^{(t-1)},\mathbf{z}_{j}^{(t-1)}, \epsilon_{ji})$,
where $\epsilon_{ji}$ is the attribute associated with the undirected edge $e_{ji}$ and $\mathbf{M}$ is a Multi-Layer Perceptron (MLP) . After aggregating all received messages, the node $x_i$ is updated by $\mathbf{\bf{z}}_{i}^{(t)} = \mathbf{U}(\mathbf{z}_{i}^{(t-1)},\sum_{j\neq i} \mathbf{m}_{j \to i}^{(t)})$, in which $\mathbf{U}$ is a Gated Recurrent Unit (GRU)  followed by an MLP  \citep{li2016gated}. After $T$ steps,  the node vectors are fed to a readout function: $
\hat p({x_i}|\mathbf{y}) = \sigma (\mathbf{R}({\bf{z}}_i^{(T)}))
$, where $\mathbf{R}$ is another MLP and $\sigma$ denotes the softmax function. The node vector is initialized as ${{\bf{z}}_i^{(0)}} = [\mathbf{y}^T\mathbf{h}_i, -\mathbf{h}_i^T\mathbf{h}_i, \sigma^2] $ and the edge attributes of $\epsilon_{ji}$ are fixed at $[\mathbf{h}_j^T  \mathbf{h}_i, \sigma^2]$.

\begin{figure}
  \centering
  \centerline{\includegraphics[width=\linewidth]{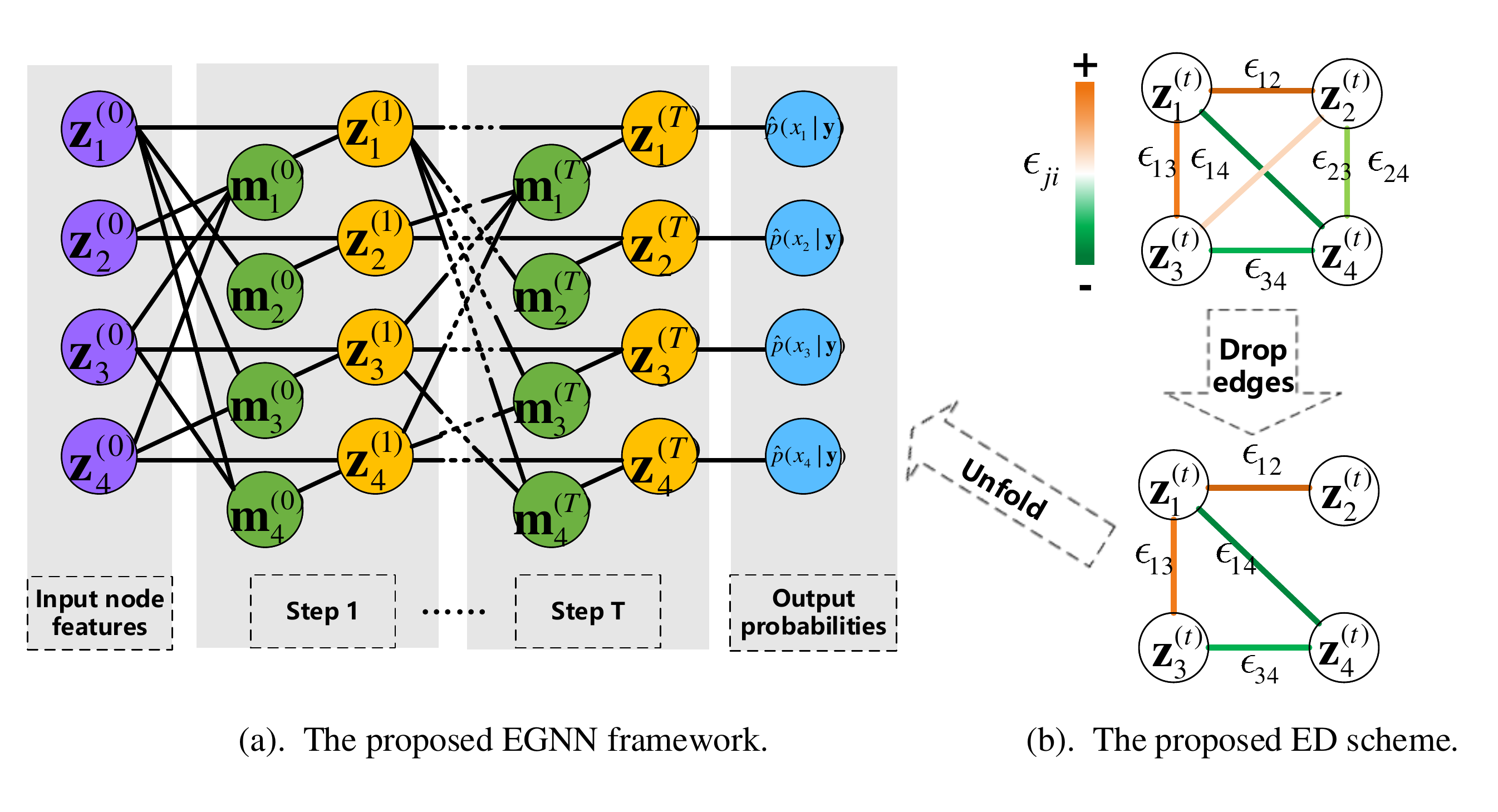}}
  \caption{The Framework Overview.}\label{fig:framework}
 \end{figure}

\section{The Proposed Edge Graph Neural Networks for MIMO Detection}\label{sec:pagestyle}

Previous works which utilize GNNs for MIMO detection have two limitations: \emph{firstly}, MLP-based message functions treat node features and edge attributes equally, and thus neglect unique edge information; \emph{secondly}, the computational cost of naive GNN models is still high because messages are passed back and forth frequently in GNNs. Therefore, we propose the EGNN and the Edge Drop (ED) scheme to address the aforementioned two issues, respectively.

\subsection{The EGNN Model}
In general, assigning different weights in neighboring messages can accelerate the learning process \citep{gong2019exploiting, velivckovic2018graph}. Moreover, neighboring information may exert either positive or negative impacts on central nodes \citep{hou2019measuring}. Take them into consideration, we propose the EGNN model as follows: 

The initial node features are mapped to a higher-dimensional space to enrich node information and enhance training performance:
\begin{align}
  {{\bf{z}}_i^{(0)}} = \mathbf{MLP}_1\left(\mathbf{y}^T\mathbf{h}_i, -\mathbf{h}_i^T\mathbf{h}_i, \sigma^2\right) \in \mathbb{R}^{S}.
  \label{node init func}
\end{align}
where $\mathbf{MLP}_1$ is an MLP layer of size ($3,S$), $S$ is the node feature dimension, and edge attributes are 
$\epsilon_{ji} = -\mathbf{h}_j^T  \mathbf{h}_i$. We then use $\epsilon_{ji}$ to measure the importance of the message from $j$ to $i$. The weighted message can be obtained by:
{\setlength\abovedisplayskip{1.5pt}
\setlength\belowdisplayskip{1.5pt}
\begin{equation}\label{proposed msg func}
\begin{aligned}
{\bf{m}}_{j \to i}^{(t)} = \epsilon_{ji} \cdot \mathbf{MLP}_2\left(\mathbf{z}_j^{(t-1)}\right),
\end{aligned}
\end{equation}
}where $\mathbf{MLP}_2$ is another MLP layer of size ($S,S$). We apply the node update function and readout function used in \cite{scotti2020graph}. The proposed EGNN framework is illustrated in Fig.\ref{fig:framework}(a). 
The preprocessing functions (\textit{i.e.} purple units) first initialize node features which are then passed to message functions (\textit{i.e.} green units) to generate messages. Then node update functions (\textit{i.e.} yellow units) collect messages to update node status. After $T$ steps, the readout functions (\textit{i.e.} blue units) provide predictions. EGNN model is optimized by minimizing the cross-entropy loss function: $L = -\sum_{i=1}^{N_t}p(x_i)log\hat{p}(x_i|\mathbf{y})$, where $p(x_i)$ is a ground-truth label vector.

\subsection{The Edge Drop scheme}
We propose the ED scheme to sparsify the fully-connected graph. Note that the edge attribute $\epsilon_{ji} = -\mathbf{h}_j^T  \mathbf{h}_i$ describes the unnormalized spacial correlation between channels of node $i$ and $j$. The higher its absolute value is, the more correlated the two channels are, and hence more weight should be imposed on them. In contrast, a close-to-zero $\epsilon_{ji}$ indicates the approximate orthogonality of two channels which need not exchange any or much information to recover the transmitted symbols.  Furthermore, naive GNN detectors require $\mathcal{O}(N_t^2)$ message-passing processes at each iteration. Therefore, the proposed ED method drops a certain amount of edges from the graph to reduce computational complexity without performance loss. The ED process is illustrated by Fig. \ref{fig:framework}(b), where the example graph has four nodes and is at first fully-connected. The red edges mean that the edge weights are positive, while the green edges mean the opposite. 
The lighter the colors are, the smaller the absolute values of the weights will become and the more trivial the edges will be such that they can be neglected ($\epsilon_{23}$ and $\epsilon_{24}$ in Fig. \ref{fig:framework}(b)).





\section{Experiments}\label{experiment}
In this section, we conduct extensive experiments to evaluate the performance of our proposed EGNN MIMO detector (with ED scheme) against existing popular methods: naive GNN and  \textit{Minimum Mean Square Error} (MMSE). The idea of MMSE is to minimize the Mean-Square Error (MSE) between the transmitted $\mathbf{x}$ and the estimated signal $\mathbf{H}^T\mathbf{y}$. Due to the lack of real-world datasets, we constructed synthetic datasets as follows: we adopted the well-known \textit{Quadrature Phase Shift Keying} (QPSK) and \textit{16-Quadrature Amplitude Modulation} (16-QAM) schemes to build the modulation alphabet $\mathcal{A}=\{s_k\}^K_{k=1}$. Then, $x_i$ on the $i$-th antenna was randomly picked from  $\mathcal{A}$ under the power constraint $\mathbb{E}\{\mathbf{x}^T\mathbf{x}\}=N_t$. 
The channel $\mathbf{H}$  was constructed from i.i.d. Rayleigh fading model, whose entries were randomly sampled from the Gaussian distribution $\mathcal{N}(0,\sigma^2)$ with $\sigma^2=1/N_r$, and $N_t = 16$ and $N_r = 32$.

\begin{figure}
  \centering
  \label{sim_QPSK}
  \begin{minipage} {\linewidth}
  \centerline{\includegraphics[width=\linewidth]{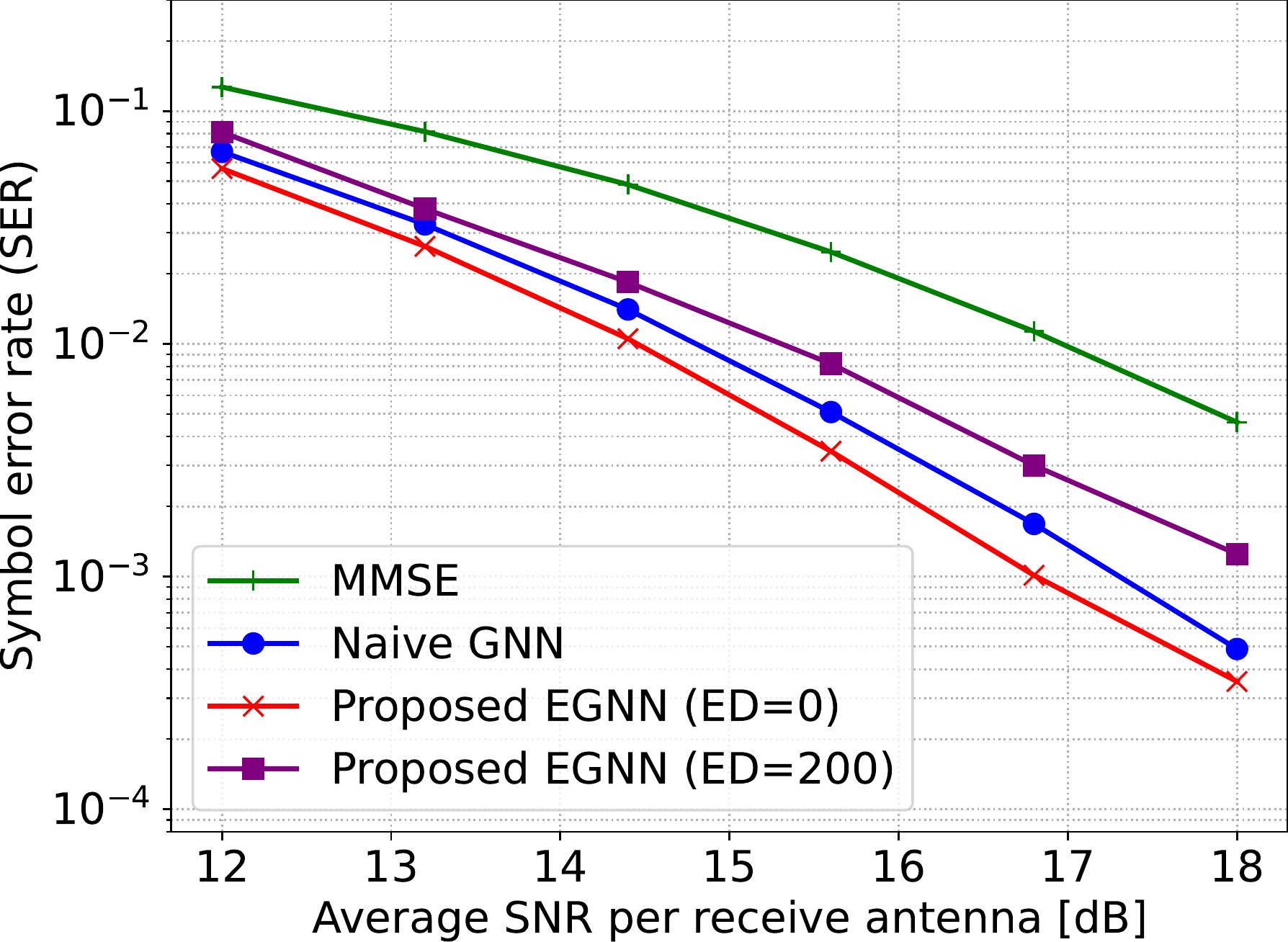}}
  \centerline{(a). 16-QAM  modulation.}
  \end{minipage}
 \vfill
 \begin{minipage}{\linewidth}
   \centerline{\includegraphics[width=\linewidth]{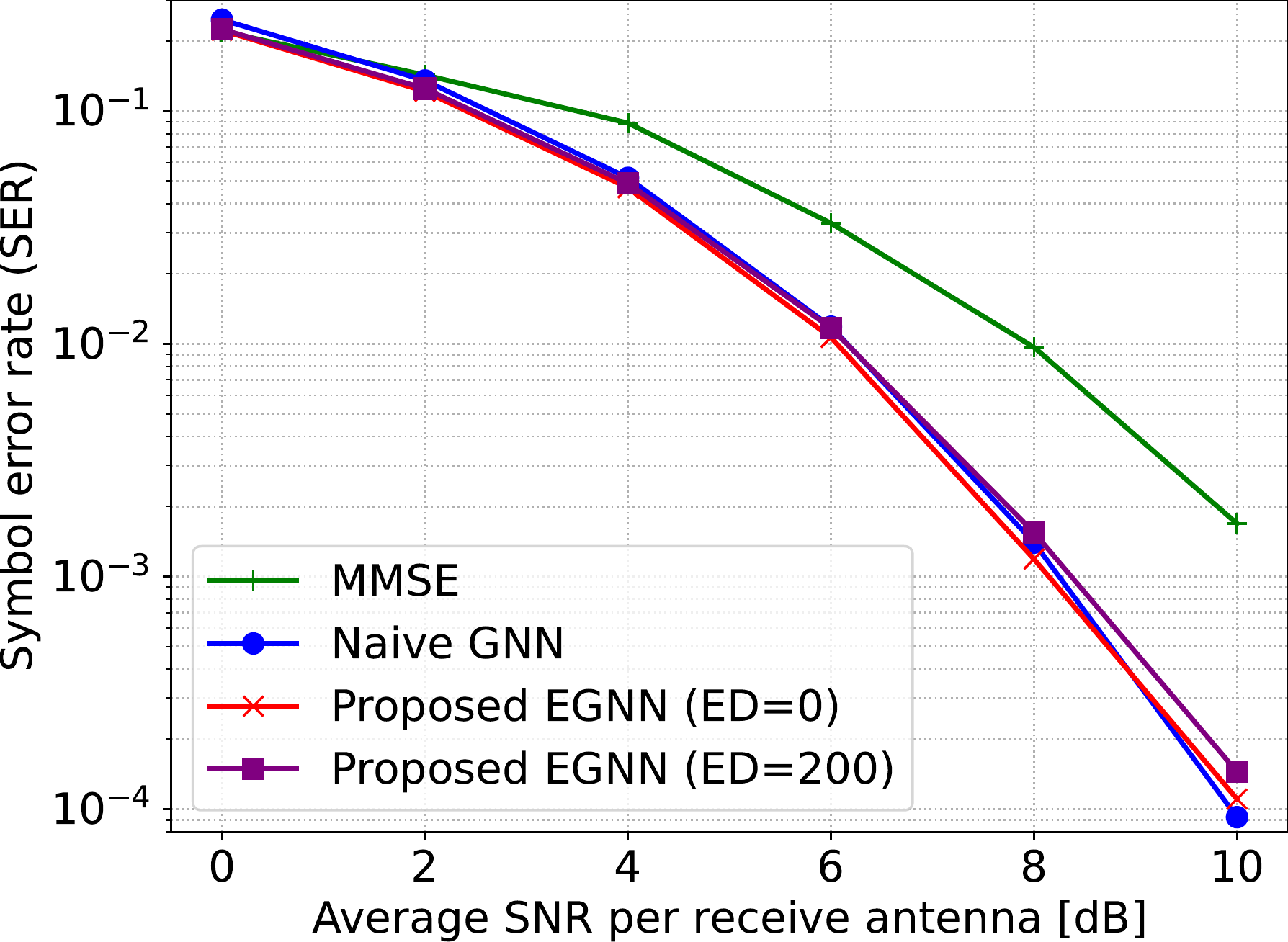}}
\centerline{(b). QPSK modulation.}
 \end{minipage}
 \caption{SER vs. SNR of different schemes where $N_t=16$, $N_r=32$.}\label{fig:ser}
\end{figure}

\begin{table}[th!]\caption{Time of training and testing an epoch for GNN-based detectors.}
\centering
\begin{tabular}{c|c|c|c}
\hline\hline
 \multirow{2}{*}{Detectors}        &  \multirow{2}{*}{Naive GNN} & \multicolumn{2}{c}{The Proposed EGNN} \\ 
 \cline{3-4}&&ED=0&ED=200\\\hline
Training time (s) / epoch & 22.10     & 17.85     & 16.75       \\\hline 
Test time (s) / epoch & 3.85     & 3.75     & 3.50       \\\hline \hline
\end{tabular}\label{tab:time}
\end{table}

We provided three GNN-based models: the naive GNN, the EGNN without dropping edges (\textit{i.e.} ED=0), and the EGNN with 200 edges dropped (\textit{i.e.} ED=200).  The numbers of training, validation, and test datasets used in our paper were 49152, 16384, and 16384, respectively. All models consisted of $T=6$ steps, and we trained them in $100$ epochs using the Adam optimizer \cite{kingma2015adam} with a learning rate of $0.0003$. Other hyperparameters were chosen as 32 (node feature dimension $S$), 128 (number of hidden neurons in the GRU module), and 64 (batch size).  Besides, experimental results of the MMSE method were averaged over 20000 Monte-Carlo experiments.  The Symbol-Error-Rate (SER) metric was utilized to compare performance. 

Fig.\;\ref{fig:ser}(a) depicts the SER performance of all methods in terms of SNR under the 16-QAM modulation scheme. It demonstrates that our proposed EGNN(ED=200)  was ahead of all other methods. To be detailed, it outperformed the Naive GNN by 0.5dB at the SER $10^{-3}$. Furthermore, our scheme with ED=200 was still superior to the MMSE method.
Fig. \ref{fig:ser}(b) demonstrates the SER results under QPSK modulation, in which our proposed EGNN detector achieved comparable performance to the state-of-the-art one. 

Table\;\ref{tab:time} shows the offline training and online test time of each epoch for GNN-based detectors. The proposed EGNN (ED=200) consumed the least offline training time and the least testing time online of each epoch, which manifested the effectiveness of our proposed ED scheme. \\
 \begin{figure}
  \centering
  \label{convergence}
  \centerline{\includegraphics[width=\linewidth]{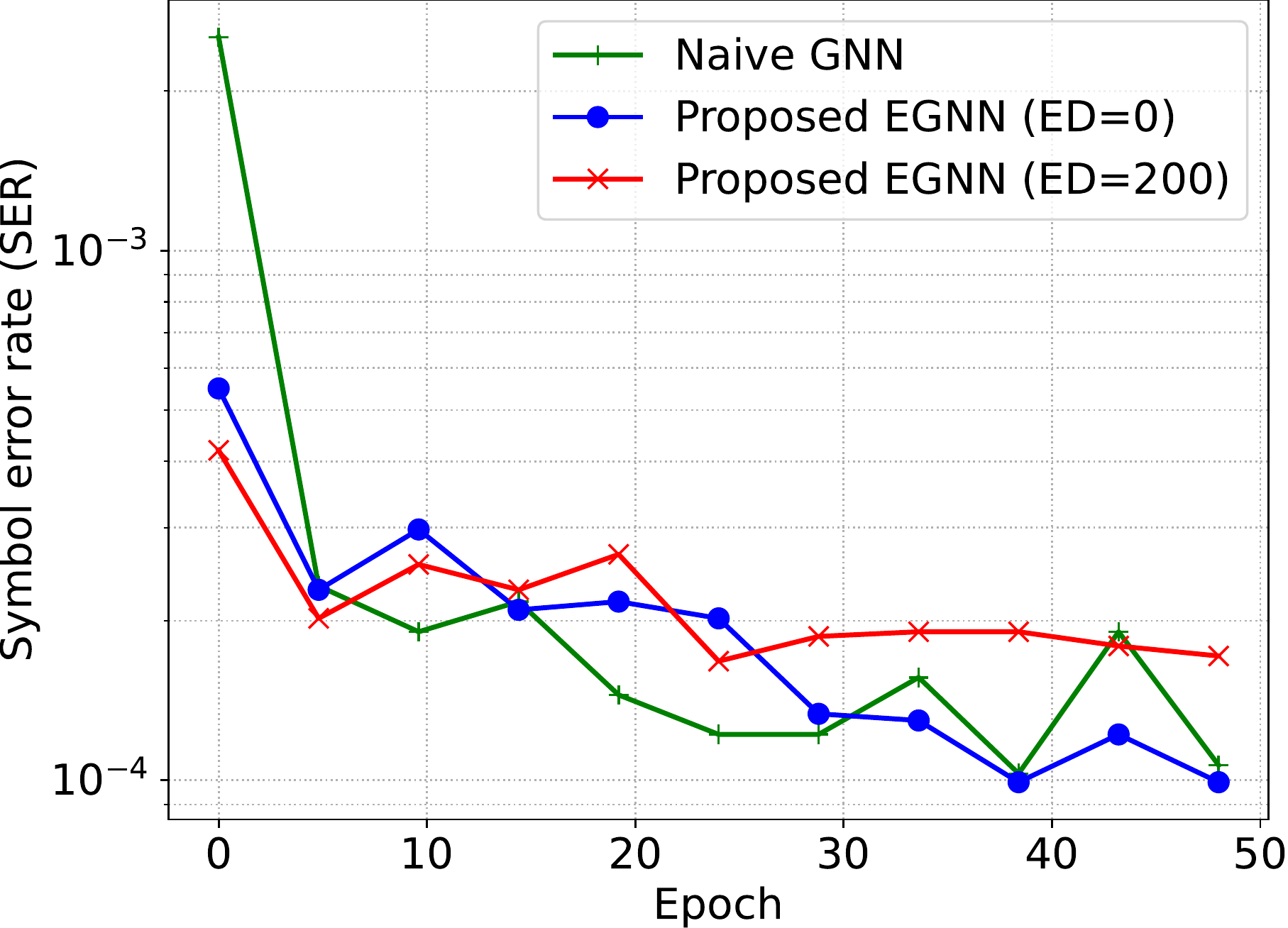}}
\caption{Convergence curves of SER under different GNN-based schemes for QPSK modulation under SNR=10dB.}\label{fig:convergence}
\end{figure}
In this experiment we further compare the convergence speed of  between GNN-based approaches.  
The SER of the first 50 epochs of the proposed EGNN (with ED) and Naive GNN detectors are reported in Fig.\;\ref{fig:convergence}, where the modulation scheme is QPSK and SNR is 10dB.
It is observed that the proposed schemes EGNN (ED=0) and  EGNN (ED=200) converged faster than Naive GNN, though they achieved close final results. 

\section{Conclusion}

In this paper, we developed an efficient Graph Neural Networks (GNN)-inspired algorithm, called Edge GNN (EGNN), to detect MIMO signals.  
We firstly proposed to compute graph edge weights by leveraging the channel correlation and then used the obtained weights as a metric to evaluate the importance of neighbors of each node. Towards fast detection, we designed an Edge Drop (ED) scheme to remove edges with a much smaller impact on message passing, where the sparsified graph was more efficient to transmit information and faster to converge. 
Experimental results demonstrate that our proposed EGNN was superior or comparable to popular MIMO detection methods using different modulation schemes and it had the least detection time compared to GNN-based approaches. 

\label{sec:Conclusion}

\vfill\pagebreak

\bibliographystyle{ACM-Reference-Format}
\bibliography{reference}

\newpage
\appendix

\end{document}